\documentstyle[seceq,mbf,wrapft]{ptptex}

\notypesetlogo  
\preprintnumber[3cm]{
KUNS-1325\\PTPTeX ver.0.8\\ August, 1997}
\title{A Dynamical Brane in the Gravitational Dual of 
${\cal N}=2$ $Sp(N)$ Superconformal Field Theory}

\author{
Shin'ichi {\sc Nojiri}\footnote{E-mail: nojiri@cc.nda.ac.jp}, 
Sergei D. {\sc Odintsov}$^{*,}$\footnote{On leave 
from Tomsk Pedagogical University, 634041 Tomsk, Russia. \\
E-mail: odintsov@ifug5.ugto.mx, odintsov@mail.tomsknet.ru}
and Sachiko {\sc Ogushi}$^{**,}$\footnote{JSPS Research Fellow. \\
E-mail: g9970503@edu.cc.ocha.ac.jp}
}

\inst{
Department of Applied Physics, 
National Defence Academy, Yokosuka 239-8686, Japan \\
$^*$Instituto de Fisica de la Universidad de Guanajuato,
Lomas del Bosque 103, Apdo. Postal E-143, 37150 Leon, Gto., 
Mexico \\
$^{**}$Department of Physics, Ochanomizu University, 
Tokyo 112-0012, Japan}

\recdate{
June 21, 2000}

\abst{
The particular model of d5 higher derivative gravity which is dual to
${\cal N}=2$ $Sp(N)$ SCFT is considered. A (perturbative) AdS black 
hole in such theory is constructed in the next-to-leading order 
of the AdS/CFT correspondence. The surface counterterms are fixed by 
the conditions required for a well-defined variational procedure 
and the finiteness of AdS space (when the brane goes to infinity).
A dynamical brane is realized at the boundary of an AdS black 
hole with a radius that is larger than the horizon radius. 
The AdS/CFT correspondence dictates the parameters of 
the gravitational dual in such a way that the dynamical brane 
(the observable universe) 
always occurs outside the horizon.
}

\begin{document}

\tolerance=5000

\def\pp{{\, \mid \hskip -1.5mm =}}
\def\cL{{\cal L}}
\def\be{\begin{equation}}
\def\ee{\end{equation}}
\def\bea{\begin{eqnarray}}
\def\eea{\end{eqnarray}}
\def\tr{{\rm tr}\, }
\def\nn{\nonumber \\}
\def\e{{\rm e}}
\def\D{{D \hskip -3mm /\,}}
\def\Box{\mbox{\fbox{\rule[0cm]{0cm}{.15cm}\ }}}

\maketitle

\makeatletter
\if 0\@prtstyle
\def\asp{0.5em} \def\bsp{0em}
\else
\def\asp{.3em} \def\bsp{0.3em}
\fi \makeatother

\section{Introduction}

In modern studies of brane-worlds, one assumes that the observable 
universe lies as a boundary in multi-dimensional bulk space, 
where gravity on the brane is trapped \cite{RS}, and bulk 
represents some AdS-like multidimensional background which 
follows from (IIB) string theory. The brane-world cosmology 
\cite{CH,cosm} (and references therein) is usually constructed 
from the (d5) AdS bulk space with the following conditions:

\noindent
a. There are few free parameters of the that must be fine-tuned 
to get the desirable properties. In most cases these parameters 
are the (negative) bulk cosmological constant and brane cosmological 
constant (brane tension).
Playing with these two parameters one has already constructed 
various brane-worlds in Einstein gravity (with surface terms).
The coefficients of higher derivative terms in the bulk 
(or brane) action may play the role of these parameters as well.

\noindent
b. As the bulk, one chooses AdS space or its product with some 
other manifold.

Clearly, the above mechanism for the realization 
of the observable brane universe is not dynamical, as 
the parameters of the theory are fine-tuned to get brane-world.
Presumably, a dynamically generated brane should be searched 
for within the AdS/CFT correspondence \cite{AdS}, where 
a warped compactification introduces the brane-world set-up. 
Indeed, one version of such a scenario that is more suitable 
in the framework of the AdS/CFT correspondence has been 
presented in Refs.\citen{sergio}, where the brane 
tension is fixed from the beginning, but the quantum CFT 
existing on the brane produces an effective brane tension.
This scenario is much more restrictive because brane-worlds 
are produced completely dynamically. Our approach follows 
this line.

 From another point of view, it is not clear what the bulk 
should be. It is expected that the topology \cite{McInnes} 
should be very important in realization of warped 
compactification in string theory in frames of the AdS/CFT 
correspondence. In a recent paper \cite{SO} we considered 
the particular model of d5 higher derivative gravity that 
possesses a Schwarzschild-anti de Sitter (S-AdS) black hole 
as an exact bulk solution. It has been shown that a 
four-dimensional brane (flat or de Sitter) 
could be realized dynamically. Hence, the observable universe 
may represent the boundary of a higher 
dimensional AdS black hole.

In the present work we generalize the mechanism of 
Ref.\citen{SO} and show that it may be realized completely 
within the AdS/CFT correspondence. The starting point 
is the particular model of d5 higher derivative (HD) 
gravity that appears as the low-energy limit of compactified 
IIB string theory. All parameters of the model are defined. 
The theory is expected to be dual to ${\cal N}=2$ SCFT 
with the gauge group $Sp(N)$ \cite{FS} (see also Ref.\citen{Sen}). 
 This duality has also been checked by comparison of holographic 
and QFT conformal anomalies in Refs.\citen{BNG} and \citen{SNhd} 
in the next-to-leading order of the large-$N$ expansion. 

The surface counterterms are added to the starting action 
in frames of the AdS/CFT correspondence. The parameters of these 
counterterms are not arbitrary. They are fixed by two conditions:

\noindent
a. The variational procedure should be well-defined.

\noindent
b. The leading divergences of a bulk AdS black hole should cancell 
(finiteness). 

In particular, the brane tension is fixed by these conditions. 

Working with such a gravitational dual of ${\cal N}=2$ SCFT, we 
construct a AdS black hole perturbatively in the 
next-to-leading order of the AdS/CFT correspondence. The 
(flat) brane is dynamically 
created as the boundary of such an AdS black hole 
and its radius, which is larger than horizon radius, is found. 
An analogous AdS-like cosmological model is also briefly discussed.


\section{AdS black hole with dynamical brane} 

Let us start from the general action of higher derivative gravity.
It is given by \footnote{The conventions for curvatures are 
defined by
\bea
&& R=g^{\mu\nu}R_{\mu\nu},\quad R_{\mu\nu}
=-\Gamma^{\lambda}_{\mu\lambda,\nu}
+\Gamma^{\lambda}_{\mu\nu,\lambda}-\Gamma^{\eta}_{\mu\lambda}
\Gamma^{\lambda}_{\nu\eta}+\Gamma^{\eta}_{\mu\nu}
\Gamma^{\lambda}_{\lambda \eta} \ ,\nn
&& R^{\lambda}_{\mu\nu\kappa}=\Gamma^{\lambda}_{\mu\kappa,\nu}
-\Gamma^{\lambda}_{\mu\nu,\kappa}+\Gamma^{\eta}_{\mu\kappa}
\Gamma^{\lambda}_{\nu\eta}-\Gamma^{\eta}_{\mu\nu}
\Gamma^{\lambda}_{\kappa \eta},\quad
\Gamma^{\eta}_{\mu\lambda}={1\over 2}g^{\eta\nu}
(g_{\mu\nu,\lambda}+g_{\lambda \nu ,\mu}-g_{\mu \lambda ,\nu})\ .
\nonumber
\eea }
\be
\label{ac0}
S=\int d^{d+1} x \sqrt{-\hat G}\left\{a \hat R^2 
+ b \hat R_{\mu\nu}\hat R^{\mu\nu}
+ c \hat R_{\mu\nu\xi\sigma}\hat R^{\mu\nu\xi\sigma}
+ {1 \over \kappa^2} \hat R - \Lambda \right\}\ .
\ee
The equations of motion derived from  the above action 
(\ref{ac0}) are 
\bea
\label{aiii}
0&=&-{1 \over 2}\hat G_{\zeta\xi}
\left\{a \hat R^{2} 
+ b \hat R_{\mu\nu}\hat R^{\mu\nu} 
+ c \hat R_{\mu\nu\rho\sigma}\hat R^{\mu\nu\rho\sigma}
+ {1 \over \kappa^2} \hat R - \Lambda \right\} \nn
&& + 2a \hat R \hat R_{\zeta\xi} 
+ 2b \hat R_{\mu\zeta}{\hat R^{\mu}}_\xi
+ 2c \hat R_{\zeta\mu\nu\rho}\hat R_\xi^{\mu\nu\rho}
+ {1 \over \kappa^2} \hat R_{\zeta\xi}\nn
&& - a\left(D_\zeta D_\xi + D_\xi D_\zeta\right) \hat R \nn
&& + b\left(\hat G_{\zeta\xi} D_\rho D_\sigma \hat R^{\rho\sigma}
 - D_\xi D_\sigma \hat R_\zeta^\sigma
 - D_\zeta D_\sigma \hat R_\xi^\sigma + \Box \hat R_{\xi\zeta} 
\right) + 4c D_\rho D_\kappa \hat R_{\zeta\ \xi}^{\ \rho\ \zeta}\ .
\eea
We choose the metric
\be
\label{met01}
ds^{2}=\hat{G_{\mu\nu}}dx^{\mu}dx^{\nu}
=-\e^{2\rho}dt^{2}+\e^{-2\rho}dr^{2}+r^{2}\sum_{i=1}^{d-1}
\left(dx^{i}\right)^2\ .
\ee
Here the 4d part of the metric is chosen to be flat only 
for simplicity. There is no problem in considering 
the de Sitter or anti-de Sitter 4d part 
of the metric. (The only difference in this case is that 
the corresponding calculations are technically a bit more 
involved.)

If we further assume $a$, $b$ and $c$ are small compared with 
${1 \over \kappa^2}$, we obtain the 
following AdS black hole solution when $d+1=5$:
\be
\label{slgn} 
\e^{2\rho} = {1\over r^2} \left\{ -\mu 
+\left(1 + {20a\kappa^2 \over 3} + {4b\kappa^2 \over 3} 
+ {2c\kappa^2 \over 3} \right)r^4
+ {2c\kappa^2\mu^2 \over r^4 } \right\}\; .
\ee
As the leading-order effect of $a$ and $b$ is simply to change 
the radius of AdS, we only consider the case $a=b=0$ in the 
following  (only Riemann curvature squared terms):
\bea
\label{ac1}
S &=& \int d^{d+1}x \sqrt{-\hat{G}}
\left\{ {1\over \kappa ^{2} }\left(\hat{R} + 12 \right)
+c\hat{R}_{\mu\nu\xi\sigma}\hat{R}^{\mu\nu\xi\sigma} 
 \right\} .
\eea
In particular, when $d+1=5$ and 
\be
\label{N2}
{1 \over \kappa^2}={N^2 \over 4\pi^2}\ ,\quad 
c={6N \over 24\cdot 16\pi^2}\ ,
\ee
the action (\ref{ac1}) appears as the low-energy theory of 
type IIB string theory on $AdS_5\times X_5$ where 
$X_5=S^5/Z_2$ \citen{FS,Sen}. This theory is believed to be dual 
to ${\cal N}=2$ theory with the gauge group $Sp(N)$.

When $a=b=0$, the perturbative solution in (\ref{slgn}) looks like 
\cite{SC} 
\be
\label{met1}
\e^{2\rho} = {1\over r^2} \left\{ -\mu 
+\left(1+{2 \over 3}\epsilon \right)r^4
+2\epsilon {\mu^2 \over r^4 } \right\},\quad \epsilon \equiv 
c\kappa^2 \; .
\ee
In the following, we ignore the terms containing 
higher powers of $\epsilon$. 
If we assume the metric to take the form in (\ref{met01}), 
the components of the Ricci tensor and Riemann tensor are given by
\bea
\hat{R}_{tt}&=& \left( \rho''+2(\rho')^{2}+{(d-1)\rho' \over r}
\right)\e^{4\rho } \ ,\nn
\hat{R}_{rr}&=& -\rho ''-2(\rho ')^{2}-{(d-1)\rho' \over r}\ , \nn
\hat{R}_{ij}&=& \left(-2r\rho' -d+2\right) \e^{2\rho} 
\delta_{ij}\ , \nn
\mbox{\rm other Ricci tensor components } &=& 0\ , \\
\hat{R}_{trtr}&=&-\hat{R}_{trrt}=-\hat{R}_{rttr}
=\hat{R}_{rtrt} \ ,\nn
&=&\e^{2\rho}\left( \rho '' +2\rho'^{2} \right)\nn
\hat{R}_{titj}&=&-\hat{R}_{ittj}=-\hat{R}_{tijt}
=\hat{R}_{itjt}\ ,\nn
&=&r\rho'\delta_{ij}\e^{4\rho}\ , \nn
\hat{R}_{rirj}&=&-\hat{R}_{rijr}=-\hat{R}_{irrj}
=\hat{R}_{irjr}\ ,\nn
&=& -r\rho '\delta_{ij}\ , \nn
\hat{R}_{ijkl}&=&-r^{2}\e^{2\rho}\left( \delta_{ik}\delta_{jl}-
\delta_{il}\delta_{jk} \right)\ , \nn
\mbox{\rm other Riemann tensor components } &=& 0\ .
\eea
The scalar curvature and the square of the Riemann tensor are
\bea
\hat R&=& \left(-2\rho ''-4(\rho')^{2}-{4(d-1)\rho' \over r}
-{(d-2)(d-1)\over r^{2} } \right)\e^{2\rho } \nn
\hat{R}_{\mu\nu\xi\sigma}\hat{R}^{\mu\nu\xi\sigma} 
&=& 4\hat{R}_{trtr}\hat{R}^{trtr} 
+4\hat{R}_{titj}\hat{R}^{titj} 
+4\hat{R}_{rirj}\hat{R}^{rirj} 
+\hat{R}_{ijkl}\hat{R}^{ijkl} \nn
&=& 4 \e^{4\rho}\left( \rho '' +2\rho'^{2} \right)^{2}
+4(d-1)r^{-2}\rho'^{2}\e^{4\rho}+4(d-1)r^{-2}\rho'^{2}\e^{4\rho}\nn
&& +2(d-1)(d-2)r^{-4}\e^{4\rho} \\
&=& 4 \e^{4\rho}\left( \rho '' +2\rho'^{2} \right)^{2}
+8(d-1)r^{-2}\rho'^{2}\e^{4\rho}
+2(d-1)(d-2)r^{-4}\e^{4\rho} . \nonumber 
\eea

We assume that there is a boundary at $r=r_0$, where the brane 
lies. Then we need to add  a boundary term, specifically, a 4 
dimensional cosmological term, in order to realize 
brane-world universe, i.e. the RS scenario for 4d gravity.\cite{RS} 
Usually such a surface term is chosen to be arbitrary.
Its fine-tuning is responsible for the creation of the brane-world.

In our scenario (within the AdS/CFT correspondence), the surface 
counterterms are not arbitrary. Such a surface term causes the 
variational principle to be well-defined and a complete 
AdS space to be finite when the brane goes to infinity. 
We take the surface terms in the following form:\cite{NOch} 
\bea
\label{ac2}
S_{b} &=& S_{b}^{(1)}+S_{b}^{(2)} \\
S_{b}^{(1)} &=& \int d^{4}x \sqrt{\hat{g}} \Bigl[
4\tilde{a}\hat{R}D_{\mu}n^{\mu} +2\tilde{b_{1}}n_{\mu}n_{\nu}
\hat{R}^{\mu \nu}D_{\sigma}n^{\sigma}+2\tilde{b_{2}}\hat{R}_{\mu\nu}
D^{\mu}n^{\nu} \nn
&& +8\tilde{c}n_{\mu}n_{\nu}
\hat{R}^{\mu\tau\nu\sigma}D_{\tau}n_{\sigma}
+{ 2\over \tilde{\kappa}^{2} } D_{\mu}n^{\mu} \Bigr]\ , \nn
S_{b}^{(2)} &=& -\eta \int d^{4}x \sqrt{\hat{g}}\ , \nonumber
\eea
where the normal vector $n^{\mu}$ and its covariant derivatives 
are given by
\bea
n^{r} &=& \e^{\rho}, \quad \mbox{\rm other components}=0 \nn
D_{r}n^{r} &=& 0, \quad D_{t}n^{t}=\e^{\rho}\rho', \quad 
D_{i}n^{j}={\e^{\rho}\over r}\delta^{j}_{i} \nn
D_{\mu}n^{\mu}&=&\e^{\rho}\rho' +{(d-1)\e^{\rho} \over r} \ .
\eea
In (\ref{ac2}), we can choose $\tilde b_1 = \tilde b_2$, 
but as we see below, it is convienient to treat them as 
independent parameters when we considers the black hole 
background as in Ref.\citen{SO}. 


When we substitute the solution (\ref{met1}) into the bulk 
action (\ref{ac1}) with $d=4$, there appears a divergence 
if there is no brane, which can be a boundary of the spacetime. 
In order to regularize the divergence, we restrict the 
integration over $r$ to be a finite region $\left(\int d^5x 
\rightarrow \int d^4x \int_0^r dr\right)$ and we assume the surface 
terms in (\ref{ac2}) appear on the boundary. 
The parameter $\eta$ in (\ref{ac2}) is determined by the 
condition that the corresponding term cancel the leading 
divergence. After the integration over $r$, we find
\be
\label{Srd}
S\sim {r^4  \over \kappa^2}\left(-2 + {20 \over 3}\epsilon
\right)\int d^4x  + o\left(r^4\right)\ ,
\ee
and the surface terms in (\ref{ac2}) behave, 
when $r$ is large, as follows:
\bea
\label{S2rd}
S_b&\sim& r^4 \left\{ -320\tilde a - 32 \tilde b_1 
 - 32 \tilde b_2 -32 \tilde c 
+ {8\over \tilde \kappa^2}\left(1 + {2 \over 3}\epsilon\right) 
\right. \nn
&& \left. -\eta\left(1 + {1 \over 3}\epsilon\right)\right\}\int d^4x  
 + o\left(r^4\right)\ .
\eea
Then we obtain
\be
\label{eta}
\eta={1 \over \kappa^2}\left(-2 + {22 \over 3}\epsilon\right)
-320\tilde a - 32 \tilde b_1 
 - 32 \tilde b_2 -32 \tilde c 
+ {8\over \tilde \kappa^2}\left(1 + {1 \over 3}\epsilon\right) \ .
\ee
The variation of the action (\ref{ac1}) and (\ref{ac2}) on the
boundary, which lies at $r=r_{0}$, gives
\bea
\delta S |_{r=r_{0}}
&=& \int d^{4} x r_{0}^{d-1}\e^{2\rho} \nn
&& \times \left[ {1 \over \kappa^{2}} \left\{
-2\delta \rho'+ \delta\rho \left(
-8\rho' -{4(d-1)\over r_{0} } \right) \right\} \right. \nn
&& +\left. c \e^{2\rho } \left\{ 8(\rho'' +2(\rho')^{2} )
(\delta \rho'+4 \rho' \delta \rho ) 
+ {16(d-1) \over r_{0}^{2} }\rho' \delta \rho  \right\} 
\right] \ , \\
\delta S_b&=&\int d^4x r_0^{d-1}\e^{\rho} \nn
&\times&\left[ \e^{3\rho}\delta\rho''\left\{\left(-8\tilde a 
 -2\tilde b_1\right)\left(\rho' + {d-1 \over r_0}\right)
 -2\tilde b_2\rho' -8\tilde{c} \rho' \right\} \right. \nn
&& + \delta\rho'\left\{\tilde a \left\{\left( -32\rho' 
 - {16(d-1) \over r_0}\right)\left(\rho' + {d-1 \over r_0}\right)
 \e^{3\rho} \right.\right. \nn
&& + 4\left( \left(-2\rho'' - 4\left(\rho'\right)^2 
- {4(d-1)\rho' \over r_0} - {(d-2)(d-1) \over r_0^2}
\right)\e^{2\rho} \right. \nn
&& \left.\left. + {(d-1)k \over r_0^2} \right)\e^\rho
\right\} \nn
&& + \tilde b_1\left\{ \left( -8\rho' 
 - {2(d-1) \over r_0}\right)\left(\rho' + {d-1 \over r_0}\right)
 \e^{3\rho} \right. \nn
&& \left. + 2 \left(-\rho'' - 2\left(\rho'\right)^2 
 - {(d-1)\rho' \over r_0} \right)\e^{3\rho} \right\} \nn
&& + \tilde b_2 \left\{ 2 \left(-\rho'' 
 - 2\left(\rho'\right)^2 - {(d-1)\rho' \over r_0} \right)\e^{3\rho} 
 \right. \nn
&& \left. - 8\left(\rho'\right)^2\e^{3\rho}
-\e^{3\rho}\rho' {2(d-1) \over r_0} 
-\e^{3\rho}{4(d-1) \over r_0^2} \right\}
 + {2\e^\rho \over \tilde \kappa^2} \nn
&& \left. +8\tilde{c}\e^{3\rho}\left(
 -6\left(\rho'\right)^2  -\rho '' 
-{(d-1)\over r_0^2 } \right)\right\} \nn
&& + \delta\rho\left\{ 4\tilde a\left\{4\e^{3\rho}
\left(-2\rho'' - 4\left(\rho'\right)^2 
 - {4(d-1)\rho' \over r_0} \right.\right.\right. \nn
&& \left.\left.\left. - {(d-2)(d-1) \over r_0^2}
\right)\left(\rho' + {d-1 \over r_0}\right) 
+ {2\e^\rho (d-1) k \over r_0^2}
\left(\rho' + {d-1 \over r_0}\right) \right\} \right. \nn
&& + 8\tilde b_1 \left(-\rho'' - 2\left(\rho'\right)^2 
 - {(d-1)\rho' \over r_0} \right)
\left(\rho' + {d-1 \over r_0}\right)\e^{3\rho} \nn
&& + 2\tilde b_2\left\{4\rho'\left(-\rho'' - 2\left(\rho'\right)^2 
 - {(d-1)\rho' \over r_0} \right)\e^{3\rho} \right.\nn
&& \left. + {4(d-1) \over r_0^3}
\left(-2r\rho' - d +2\right)\e^{3\rho} \right. \nn
&& \left. + {2k(d-1) \over r^3}\right\} 
+ {4\e^\rho \over \tilde\kappa^2} 
\left(\rho' + {d-1 \over r_0}\right) \nn
&& \left.\left. -32\tilde{c}\e^{3\rho}\left(
(\rho''+2\rho'^{2})\rho' +{\rho'(d-1)\over r_0^2 } \right)
 -\eta \right\}\right]\ .
\eea
To satisfy the condition that the variational principle
in the theory under discussion be well-defined, the coefficients 
of $\delta \rho''$ and $\delta \rho'$ must 
vanish.  From the condition that 
the coefficient of $\delta \rho''$ vanishes, we obtain 
\bea
\label{tild}
\tilde{b}_1 = -4\tilde{a}, \quad \tilde{b}_2= -4\tilde{c}\;\; .
\eea
Then, substituting the solution (\ref{met1}) and the condition 
(\ref{tild}), we find (putting $d=4$)
\bea
\label{vr2}
\lefteqn{\delta S + \delta S_b} \nn
&=& \int d^4 x r_0^3\Bigl[ \delta\rho'
\Bigl( 2\left({1 \over \tilde{\kappa}^2 }-{1 \over \kappa^2 }
\right)+8c\left( -3r_0^{-4}\mu +1 \right) \nn
&& -12 \tilde{a}(8r_0^{-4} \mu + 12 (1-\mu r_0^{-4}))\nn
&& + 24 \tilde{c}(3+r_0^{-4}\mu ) \Bigr)\e^{2\rho}  \nn
&& +\delta\rho \Bigl( -{4 \over \kappa^2 }
\left\{ -\mu r_0^{-3} + 5 \left(1+{2\over 3}\epsilon \right)r_0
 -6 \epsilon\mu^2 r_0^{-7} \right\} \nn
&&+{4\over \tilde{\kappa}^2}
\left\{ -2r_0^{-3}\mu + 4 \left(1+{2\over 3}\epsilon \right)r_0
 \right\}\nn
&&+ 16c\left\{ -9\mu^2 r_0^{-7}-4r_0^{-3}\mu+ 5r_0 \right\} \nn
&& - 48\tilde{a}\left\{ -8r_0^{-3}\mu + 16 r_0 \right\} \nn
&& + 96\tilde{c}\left\{ 2 \mu^2 r_0^{-7} -2 \mu r_0^{-3}
+ 4r_0 \right\}\nn
&& -\eta{\sqrt{-\mu+r_0^4} \over r_0}
\left\{1+ {\epsilon r_0^{4} \over 3\left(-\mu +r_0^4\right)} 
 -{\epsilon\mu^2 \over r_0^4\left(-\mu +r_0^4\right)}
\right\} \Bigr) \Bigr]\ .
\eea
The condition that the coefficient of $\delta \rho '$ vanishes
gives $\tilde{a}$ and $\tilde{c}$ as
\bea
\label{tild2}
\tilde{a} &=& {1\over 144}\left( 40c
+\left( {1\over \tilde{\kappa}^2 }
 -{1\over \kappa^2 }\right)\right) \ ,\nn
\tilde{c} &=& {4 \over 9}c-{1 \over 72}
\left( {1\over \tilde{\kappa}^2 }-{1\over \kappa^2 }\right)\: .
\eea
Using the above expressions for $\tilde{a}$ and $\tilde{c}$, 
and as well as the relation $c={\epsilon \over \kappa ^2}$,
the condition that the coefficient of $\delta \rho$ 
vanishes leads to
\bea
\label{bdryeq}
0&=&F(r_0) \nn
&\equiv&{1\over \tilde{\kappa}^2 }\left\{ -{8\over 3}\mu^2 r_0^{-7}
-{8\over 3}\mu r_0^{-3}+{16\over 3}(1+2\epsilon)r_0
\right\} \nn
&& +{1\over \kappa ^2}\left\{ -{8\over 3}
(13\epsilon -1)\mu ^2 r_0^{-7}
-{4\over 3}(32\epsilon+1)\mu r_0^{-3} 
+\left( 24 \epsilon-{28\over 3}\right)r_0 \right\} \nn
&& -\eta{\sqrt{-\mu+r_0^4} \over r_0}
\left\{1+ {\epsilon r_0^{4} \over 3\left(-\mu +r_0^4\right)} 
 -{\epsilon\mu^2 \over r_0^4\left(-\mu +r_0^4\right)}
\right\} \ .
\eea
Using (\ref{eta}), (\ref{tild}) and (\ref{tild2}), we find that 
$\eta$ in (\ref{bdryeq}) has the following form:
\be
\label{eta2}
\eta= {2 \over 3\kappa^2}\left(1-5\epsilon\right)
+ {16 \over 3\tilde \kappa^2}\left(1 
+ {1 \over 2}\epsilon\right)\ .
\ee
Hence, the coefficients of the surface counterterms in the 
AdS/CFT correspondence are now fixed. 

Equation (\ref{bdryeq}) can be regarded as the equation to determine 
$r_0$, i.e., the position of the brane. 
When $r_0\rightarrow \infty$, $F(r_0)$ behaves as 
\bea
\label{Fl}
F(r_0) &\sim&\left\{{1\over \tilde{\kappa}^2 }
\left({16\over 3} + {32 \over 3}\epsilon\right)
+{1\over \kappa ^2}\left( -{28\over 3}+24 \epsilon\right)
 -\eta\left(1 + {\epsilon \over 3}\right)\right\} r_0 \nn
&\sim& - {10 \over \kappa^2} + {\cal O}(\epsilon) < 0\ . 
\eea
On the other hand, when $r_0 \rightarrow \mu^{1 \over 4}$, 
$F(r_0)$ behaves as 
\be
\label{Fh}
F(r_0)\sim{2 \over 3}{\eta\epsilon\mu \over 
\sqrt{-\mu + r_0^4}}\ .
\ee
In case of the string theory dual to the ${\cal N}=2$ 
theory with the gauge group $Sp(N)$ in (\ref{N2}), 
$c$ and, therefore, $\epsilon$ are positive. Combining 
(\ref{Fl}) and (\ref{Fh}), we find that 
there is a solution  $r_0$ satisfying the brane equation 
(\ref{bdryeq}) in the ${\cal N}=2$ SCFT case. 
Equations (\ref{Fl}) and (\ref{Fh}) imply that $r_0^4 - \mu
={\cal O}\left(\epsilon^2\right)$. Then, assuming 
\be
\label{sl1}
r_0^4 =\mu + \alpha^2 \epsilon^2 
+ {\cal O}\left(\epsilon^3\right) \quad (\alpha>0) 
\ee
and substituting (\ref{eta2}) and (\ref{sl1}) into 
(\ref{bdryeq}), we find
\be
\label{sl2}
\alpha={2 \over 3}\cdot {1 + {\tilde \kappa^2 \over 8\kappa^2} 
\over 1 + {13\tilde \kappa^2 \over 8\kappa^2}} 
+ {\cal O}\left(\epsilon\right)\ .
\ee
In particular, if we choose $\tilde \kappa^2 = \kappa^2$, as in the 
original Gibbons-Hawking term,\cite{3} we obtain 
\be
\label{sl3}
\alpha={2 \over 7}\ ,
\ee
which yields 
\be
\label{sl4}
r_0^4=\mu + \left({2 \over 7}\epsilon\right)^2 
+ {\cal O}\left(\epsilon^3\right)\ .
\ee
We should note that the solution for $r_0$ gioven by (\ref{sl2}) 
[with (\ref{sl1})] or (\ref{sl4}) represents a larger value 
than the unperturbative horizon which lies at $r=\mu^{1 \over 4}$ 
(in terms of mass of the AdS black hole under consideration). 
As we see, the  $c$ correction makes 
the radius of the horizon smaller if $c$ or $\epsilon$ is positive. 
Then, the brane always exists outside the horizon. 
In other words, AdS/CFT duality predicts the correct signs of 
the gravitational action in such a way that the observable 
universe is realized as the brane outside the multi-dimensional 
black hole horizon. The whole evolution of theuniverse could 
occur within less than one period of the black hole time.

Let us consider the thermodynamic quantities. In the solution 
given by (\ref{met1}), the radius $r_h$ of the horizon and 
the temperature $T$ are given by
\be
\label{sp3}
r_h\equiv \mu^{1 \over 4}\left(1 - {2 \over 3}\epsilon \right)\ ,
\quad T={\mu^{1 \over 4} \over \pi }\left(1 - 2\epsilon\right)
={\mu^{1 \over 4} \over \pi }\left(1 - {1 \over 8N}\right)\ .
\ee
After Wick-rotating the time variables by $t\rightarrow i\tau$, 
the free energy ${\cal F}$ can be obtained from the action $S$ in 
(\ref{ac0}) with $a=b=0$, where the classical solution 
is substituted. We find
\be
\label{F1}
{\cal F}={1 \over T}S\ .
\ee
Using (\ref{aiii}) with $a=b=0$, (\ref{N2}) and (\ref{met1}), 
we find
\bea
\label{sp5}
S&=&{N^2 \over 4\pi^2}\int d^5x \sqrt g \left\{ 8 
- {2\epsilon \over 3}\left(40 + {72 \mu^2 \over r^8}
\right)\right\} \nn
&=& {N^2V_3 \over 4\pi^2 T}\int_{r_h}^\infty dr r^3 
\left\{ 8 - {2\epsilon \over 3}\left(40 + {72 \mu^2 \over r^8}
\right)\right\}\ .
\eea
Here $V_3$ is the volume of 3d flat space, and 
we have assumed that $\tau$ has a period of ${1 \over T}$. 
The expression of $S$ 
contains the divergence coming from large $r$. In order to 
subtract the divergence, we 
regularize $S$ in (\ref{sp5}) by cutting off the integral at 
a large radius $r_{\rm max}$ and subtracting the solution 
with $\mu=0$. This yields
\bea
\label{sp7}
S_{\rm reg}&=&{N^2V_3 \over 4\pi^2 T}\left(\int_{r_h}^\infty dr r^3 
\left\{ 8 - {2\epsilon \over 3}\left(40 + {72 \mu^2 \over r^8}
\right)\right\}\right. \nn 
&& \left. - \e^{\rho(r=r_{\rm max}) - \rho(r=r_{\rm max};\mu=0)}
\int_0^{r_{\rm max}} dr r^3\right)
\left\{ 8 - {80\epsilon \over 3}\right\}\ .
\eea
The factor $\e^{\rho(r=r_{\rm max}) - \rho(r=r_{\rm max};\mu=0)}$ 
is chosen so that the value of the proper 
length of the circle which corresponds to the period ${1 \over T}$ 
in Euclidean time at $r=r_{\rm max}$ for the two 
solutions coincide. Then we find
\be
\label{sp8}
{\cal F}=-{N^2V_3\left(\pi T\right)^4 \over 4\pi^2 }\left(1 
+ {3 \over 4N}\right)\ .
\ee
The entropy ${\cal S}$ and the mass (energy) $E$ are given 
by
\bea
\label{sp9}
{\cal S}&=&-{d{\cal F} \over dT}={N^2V_3\left(\pi T\right)^4  
\over \pi^2 T }\left(1 + {3 \over 4N}\right) \ ,\nn
E&=&{\cal F}+T{\cal S}={3N^2V_3\left(\pi T\right)^4  \over 4\pi^2 }
\left(1 + {3 \over 4N}\right)\ .
\eea
Hence, as is expected, the presence of a non-trivial boundary 
does not influence the black hole thermodynamics,\cite{SC} 
which provides the corresponding description for dual SCFT 
at finite temperature in the next-to-leading order 
of the large-$N$ expansion.

\section{Discussion}

As we demonstrated with the example of the gravitational dual 
(HD gravity) of SCFT with two supersymmetries, the dynamical 
brane (the observable universe) may occur as the boundary of 
a d5 AdS black hole in the next-to-leading order of 
AdS/CFT correspondence. The coefficients of the surface 
counterterms are consistently fixed within AdS/CFT 
correspondence; they are not fine-tuned by the condition 
of the existence of the brane (which is the 
usual case in brane-world scenarios). Moreover, the signs of 
coefficients (predicted by AdS/CFT) of HD 
gravity are such that the brane radius is larger than 
the horizon radius. In other words, the observable universe 
may be realized as the boundary of a multi-dimensional AdS 
black hole but outside of the horison. 
It could be interesting to further develop
the details of such scenario.

Few remarks are in order.
First, inside the horizon $r<r_h$, if we rename $r$ as $t$ and $t$ 
as $r$, we obtain a metric corresponding to an AdS-like cosmological 
model:
\bea
\label{cBHcos1}
ds^{2}&=&-\e^{2\rho}dt^{2}+\e^{-2\rho}dr^{2}+r^{2}\sum_{i=1}^{d-1}
\left(dx^{i}\right)^2\ , \nn
\e^{-2\rho} &=& -{1\over t^2} \left\{ -\mu 
+\left(1+{2 \over 3}\epsilon \right)t^4
+2\epsilon {\mu^2 \over t^4 } \right\} \ .
\eea
The leading-order behavior of the curvature here is the same as in 
the black hole case with $c=0$, and we find that there is a 
curvature singularity at $t=0$:
\be
\label{cbhA1}
\hat R_{\mu\nu\xi\sigma}\hat R^{\mu\nu\xi\sigma}
= 40 + {72 \mu^2 \over t^8}\ .
\ee
Also there is a horizon at $t=r_h$. 
The singularity could be regarded as a kind of big bang.
The topology of the spatial 
part is $S_1\times R_3$ if we impose a periodic boundary 
condition on $r$ and it is $R\times R_3$ if we do not. 
Here $R_3$ corresponds to the coordinates $x^i$. 

We can consider the free energy analogue of ${\cal F}^{\rm cos}$ 
when $c=0$ as follows:
\bea
\label{cBHcos2}
T{\cal F}^{\rm cos}= {8N^2V_3 \over 4\pi^2 T}\int_0^{r_h} dt\, t^3 
= {2N^2V_3 \left(\pi T\right)^4 \over 4\pi^2 T}\ .
\eea
If $c\neq 0$, however, the free energy diverges, 
due to the singularity of $\hat R_{\mu\nu\xi\sigma}
\hat R^{\mu\nu\xi\sigma}$ in (\ref{cbhA1}) when 
$t\rightarrow 0$. 
This simple consideration suggestssome duality between 
black hole solutions and cosmological solutions in 
multidimensional HD gravity. As there exists a 
well-developed technique to study  cosmological models,
the above trick may be useful in the investigation of 
black hole interiors.

\section*{Acknowledgements}

The work of S.O. has been supported in part by the Japan Society 
for the Promotion of Science and that of S.D.O. by CONACyT (CP, 
Ref.990356 and Grant 28454E).

\end{document}